# Collaborative residual learners for automatic icd10 prediction using prescribed medications

Yassien Shaalan[1],PhD, Alexander Dokumentov[1], PhD, Piyapong Khumrin[2], MD,PhD, Krit Khwanngern[3], MD, Anawat Wisetborisut[4], MD, Thanakom Hatsadeang[5], Nattapat Karaket[6], Witthawin Achariyaviriya[7], Sansanee Auephanwiriyakul[8], Nipon Theera-Umpon[9], PhD, Terence Siganakis[1], MSc

**Abstract.** Clinical coding is an administrative process that involves the translation of diagnostic data from episodes of care into a standard code format such as ICD10. It has many critical applications such as billing and aetiology research. The automation of clinical coding is very challenging due to data sparsity, low interoperability of digital health systems, complexity of real-life diagnosis coupled with the huge size of ICD10 code space. Related work suffer from low applicability due to reliance on many data sources, inefficient modelling and less generalizable solutions. We propose a novel collaborative residual learning based model to automatically predict ICD10 codes employing only prescriptions data. Extensive experiments were performed on two real-world clinical datasets (outpatient & inpatient) from Maharaj Nakorn Chiang Mai Hospital with real case-mix distributions. We obtain multi-label classification accuracy of 0.71 and 0.57 of average precision, 0.57 and 0.38 of F1-score and 0.73 and 0.44 of accuracy in predicting principal diagnosis for inpatient and outpatient datasets respectively.

## 1. Introduction

Clinical coding is a demanding job requiring the knowledge of medical & anatomical terminologies, health data standards, information systems and classification conventions. Clinical coders translate all clinical information (e.g. discharge summaries, pathology tests, pharmacy orders) to standard code formats such as ICD10 to represents diagnosis for easier processing[1]. It has many real-world applications such as billing & auditing[2-3] and aetiology research and analysis[4,5,6].

One of the challenges of clinical coding is miscoding. The average human coding accuracy was found to be between 70-75%[7-8]. Errors can result in severe loses of millions of dollars in underpayments[3,8,9]. Moreover, miscoding can result in losing track of epidemic trends[5]. The low coding throughput and shortage of staff add further challenges to manual coding efforts[7]. All these challenges push toward the automation of clinical coding to narrow down this gap.

Related work in literature can be studied from multiple angles. First, the problem was casted as multi-class classification[10], multi-label classification[11,12] and ranking[13]. Second, various inputs sources were studied such as discharge summaries[14-17], radiology reports[18,19], medical imaging[21] and combined sources[11,21]. Third, multiple modelling mechanisms were utilized such as rule-based models[14], supervised machine learning models[15,16,17,22] and unsupervised models[23]. To conquer the problem complexity, deep learning solutions were proposed based on recurrent, convolutional and attention based neural networks(NN)[15,17,18,21].

Our study is motivated by a few observations that highlights gaps in previous work. First, the high dependency on many data sources leading to low applicability in real-life applications. As a result less attention was given to effectively distill knowledge from individual complex data sources. Second, the scale of data used in previously proposed models is limited (frequent codes only) which does not reflect real-life case-mix complexities. Third, the high dimensionality of ICD10 code space formulates the problem as extreme multi-labeling increasing the difficulty of classification due to problems such as the long tail problem as shown in Figure 1. In this case a large fraction of labels have a very small representation in training sets which disallow learning accurate classifiers.

[1] Growing Data yassein, alexander, terence @growingdata.com.au
[2] BioMedical Informatics center, Faculty of Medicine, Chiang Mai University
[3] Department of Surgery, Faculty of Medicine, Chiang Mai University
[4] Department of Family Medicine, Faculty of Medicine, Chiang Mai University
[5] Department of Computer Engineering, Faculty of Engineering, Chiang Mai University
[6] Department of Electrical Engineering, Faculty of Engineering, Chiang Mai University
[7] Biomedical Engineering Institute, Chiang Mai University
8 Maharaj Nakhon Chiang Mai hospital, Chiang Mai, Thailand
9 IT department of Maharaj Nakhon Chiang Mai hospital

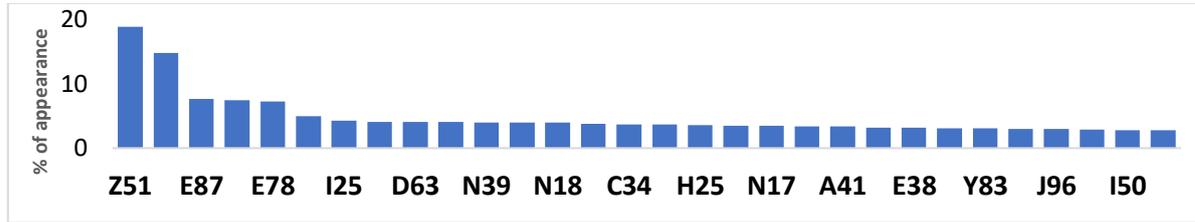

**Figure 1.** Top 30 ICD10 codes appearance frequency in inpatient dataset

In this paper, a novel deep learning model is presented based on collaborative residual learners to effectively predict ICD10 codes using prescribed medications. Towards this goal, the following contributions were made:
- The full exploitation of medication-to-diagnosis association on real case-mix distributions.
- The introduction of a novel residual learning based model to predict principal diagnosis along with comorbidities per episode of care.

To the best of our knowledge, none of existing research on medical coding fully exploits the setting of extreme multi-label classification on real-life scale while employing a single input source. Empirical and quantitative evaluation on two real-world datasets show the great potential of the proposed model in dealing with the aforementioned challenges.

## 2. Methodology

The problem can be formally defined as extreme multi-label classification problem where the output is a subset of relevant ICD-10 codes per case chosen from an extremely large set of target labels. First, the reasoning behind the modeling process (residual learning) is discussed addressing previous challenges. Then, the model structure is revealed and how it is built utilizing residual blocks.

*2.1 Residual Learning*

Deep learning has made revolutionary advances for many applications through learning complex representations to capture hidden patterns in data. To address data complexity, more layers were added to networks to increase their capacity to learn deeper knowledge. However, the high sparsity in our data led to the vanishing gradient problem obstructing the learning process from reaching a global minimum leading[24]. Traditionally in designing NNs, each layer feeds the following layer in the architecture. Residual networks (ResNets) are a special type of NNs utilizing skip connections that act as shortcuts to jump over some layers in the architecture[25]. Each layer feeds the next layer along with layers k-hops away as shown in Figure 2. ResNets usually comprise of one or more residual blocks (contain elements of nonlinearities ReLU) added in series.

Adopting this architecture has many benefits, one is to avoid vanishing gradients. Reusing activations from previous layers allows the affected layers -from vanishing gradients- to iteratively learn their weights efficiently. Consequently, the network gradually restores the skipped layers by exploring the feature space productively. Moreover, with skipping the network becomes simpler leading to faster learning process. Conversely, without skipping, the network is more vulnerable to perturbations from unfruitful deeper exploration to the feature space.

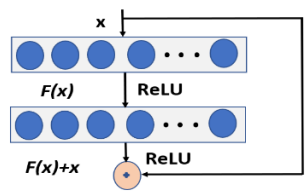

**Figure 2.** Residual neural network building block

*2.2 Collaborative Residual Learners*

The high complexity and non-linearity of medication-disease associations make deep learning a natural choice. It was long believed that the deeper the network, the more accurate it becomes[26]. Lately, such belief was defied empirically showing the training and testing error may increase after certain depths[25]. Very deep networks become more complex, slower, and more sensitive to bias and overfitting. A similar effect was experienced with this problem, and even without going very deep (only 5 layers). We witnessed the saturation and later the degradation in performance.

In this light, residual structure is adopted as the main building block. In conventional NNs, the layers directly learn the true output H(x), whereas residual networks learn the residual R(x).

$$R(x) = \text{Output} - \text{Input} = H(x) - x$$

Learning the residual of output and input was found easier than the input alone[25]. In Figure 3, a graphical representation the proposed network structure (CollabRes) is presented. Inspired by ensemble modeling, train a set of 4 individual residual networks are trained in parallel, each with two hidden layers and one drop-out with different rates. The rationale behind this is that training sub-networks with different rates and based on how the error flows backwards in the network allows for learning different information levels from taking separate paths.

For each residual block, the forward propagation through the activation function for a weight matrix $W^{l-1,l}$ connecting layer *l-1* to *l*, and a weight matrix $W^{l-2,l}$ connecting layer *l-2* to *l* is shown as follows:

$$a^\ell := \mathbf{g}(W^{\ell-1,\ell} \cdot a^{\ell-1} + b^\ell + W^{\ell-2,\ell} \cdot a^{\ell-2})$$
$$:= \mathbf{g}(Z^\ell + W^{\ell-2,\ell} \cdot a^{\ell-2})$$

where **a** is activation output and **g** is activation function.

Next, the learned knowledge is combined from the multiple blocks via concatenation. To consolidate the shared knowledge, one more residual block is added on top to incorporate original input with obtained result. This way, faster learning is assured along with the propagation of larger gradients to initial layers. Consequently, one's network dynamically realizes the importance of each layer and pathway through the assessment of the shared knowledge from each towards one's prediction goal.

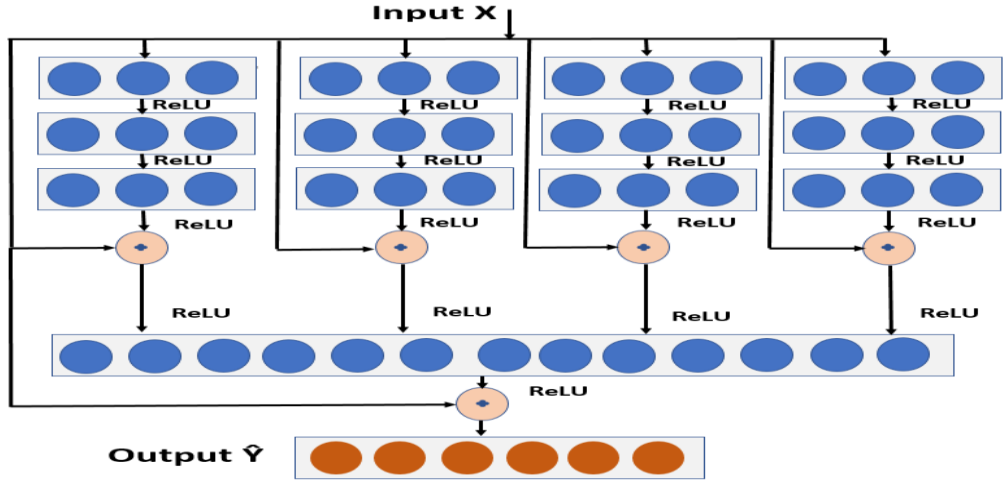

**Figure 3.** CollabRes model structure

## 3. Experimental *Setup*

*3.1 Dataset Description & Pre-Processing*

**Table 1.** Datasets high level statistics

| Dataset | #Records | #Unique Medications | Avg#Meds/case | Avg#codes/case | Max #codes /case | #Unique codes | #Unique combinations of codes |
|---|---|---|---|---|---|---|---|
| Inpatient | 588,932 | 5M | 13.47 | 3.43 | 52 | 12,382 | 250k |
| Outpatient | 475,508 | 3M | 3.22 | 1.62 | 12 | 9,970 | 80k |

The data is recorded for the period from 2006 to 2019 and Table 1 highlights the overall statistics for both datasets. To capture subtle variations in medication-disease associations, prescribed medications with a different doses are considered different. Data is filtered from cancelled prescriptions and very rare cases with less than 3 instances. Medication codes are transformed to multi-label binarized sparse vectors. Prediction of target ICD10 codes are limited first 3 letters (reflecting disease category).

*3.2 Evaluation Measures*

Assessing the performance of multi-label classification is not strictly limited to the exact number of correct label sets, but also which partial combinations are correct. The results are inspected from multiple angles to understand error types including under-coding and over-coding. Evaluation measures of Average precision, Coverage error, Ranking loss, F1 score, Jaccard similarity and Accuracy (only for primary diagnosis) are adopted[27].

*3.3 Experimental Setting & Baselines*

To reproduce one's experiments, all data were stratified and split into (train-dev-test) sets of (70%-10%-20%). Adam optimizer was employed for training on batch size of 2048 for 100 epochs with early stopping on validation accuracy of 10 iterations. The sparse input dimensions after pre-processing are 4986 and 3008 for inpatient and outpatient respectively. Keras Python library was employed for implementation. Experiments were performed on a 2 GPU (GTX 1060 6GB) machine.

Table 1 lists the baselines used to evaluate the performance accuracy. Comparisons were made against networks of varied depths (1 to 5 layers) along with a simple residual network structure using 2 layers similar to the architecture in Figure 2.

**Table 1.** Descriptions of baseline models

| Model | Description |
|---|---|
| M1 | One hidden layer (600 units with ReLU activation) |
| M2 | One hidden layer (600 units ReLU) + Dropout layer (DL) with rate 0.35 |
| M3 | Two hidden layers (600&400 units with ReLU) |
| M4 | Two hidden layers (600&400 units with ReLU) + DL with rate 0.35 |
| M5 | Three hidden layers (600, 400& 250 with ReLU) |
| M6 | Three hidden layers (600, 400& 250 with ReLU) + DL with rate 0.35 |
| M7 | Five hidden layers (600, 400, 250, 200 & 150 with ReLU) + DL with rate 0.35 |
| M8 | Two-layer residual model (600&400 units ReLU) + DL with rate 0.35 |
| CollabRes | Our proposed collaborative residual learners model |

*4. Results & Discussion*

*4.1 ICD10 Multi-label Classification*

The experimental results to empirically evaluate CollabRes's performance are shown in Table 2. A gradual increase can be seen in performance for models M1 to M5. This is directly proportion to the increase in the number of hidden layers which makes sense. However, performance stabilize for M6 and then degrade for M7 with highest model complexity. M8, the simple residual network architecture shows an improvement of 5-6% than traditional models. Yet, collaborative residual blocks introduced by CollabRes shows an extra 2-3% increase in performance and 6-7% compared to M6. In real-life over-coding is a critical concern as the number of codes to predict is not given in prior. Less over-coding errors were witnessed in CollabRes predictions suggesting that the model is more cautious and learns from shared mistakes. The same trend in performance is noticed for outpatients in Table 3. However, the figures are lower as less codes are assigned per case on average, so mistakes propagate more. Add to that no historical data is incorporated to uncover follow-up cases currently treating comorbidities.

**Table 2.** ICD10 Classification accuracy for inpatient dataset

| Model | Average Precision | Ranking Loss | Coverage Error | Jaccard Similarity | F1 | Accuracy (Primary Diagnosis) |
|---|---|---|---|---|---|---|
| M1 | 0.69 | 1.35 | 20.92 | 0.43 | 0.49 | 0.63 |
| M2 | 0.69 | 1.42 | 21.73 | 0.42 | 0.48 | 0.62 |
| M3 | 0.70 | 1.04 | 16.86 | 0.45 | 0.51 | 0.66 |
| M4 | 0.70 | 1.05 | 16.84 | 0.44 | 0.50 | 0.64 |
| M5 | 0.69 | 1.27 | 20.52 | 0.45 | 0.51 | 0.66 |
| M6 | 0.69 | 1.25 | 20.51 | 0.45 | 0.51 | 0.66 |
| M7 | 0.65 | 2.16 | 32.53 | 0.43 | 0.49 | 0.64 |
| M8 | 0.70 | 1.13 | 19.30 | 0.48 | 0.55 | 0.70 |
| CollabRes | 0.71 | 1.05 | 17.15 | 0.50 | 0.57 | 0.73 |

**Table 3.** ICD10 Classification accuracy for outpatient dataset

| Model | Average Precision | Ranking Loss | Coverage Error | Jaccard Similarity | F1 | Accuracy (Primary Diagnosis) |
|---|---|---|---|---|---|---|
| M1 | 0.54 | 2.63 | 35.39 | 0.27 | 0.28 | 0.33 |
| M2 | 0.54 | 2.65 | 35.62 | 0.26 | 0.27 | 0.31 |
| M3 | 0.55 | 2.30 | 30.10 | 0.28 | 0.30 | 0.34 |
| M4 | 0.55 | 2.35 | 30.21 | 0.27 | 0.29 | 0.33 |
| M5 | 0.55 | 2.41 | 32.68 | 0.29 | 0.30 | 0.35 |
| M6 | 0.55 | 2.40 | 32.67 | 0.29 | 0.30 | 0.35 |
| M7 | 0.54 | 2.84 | 38.18 | 0.28 | 0.30 | 0.35 |
| M8 | 0.56 | 2.19 | 30.17 | 0.33 | 0.36 | 0.41 |
| CollabRes | 0.57 | 2.14 | 28.8 | 0.35 | 0.38 | 0.44 |

Top 5 disease categories ranked by accuracy are shown in Table 4. The number of records for the top 5 categories of outpatient is ~45% of the dataset and yet average accuracy of 67% can be seen. For inpatient, the performance of one's model in detecting neoplasm related diseases is 81% (of ~30% of data).

**Table 4.** Top 5 diagnosis categories prediction accuracy

| Outpatient Dataset | | Inpatient Dataset | |
|---|---|---|---|
| Diagnosis Category | Accuracy | Diagnosis Category | Accuracy |
| Endocrine, nutritional & metabolic diseases | 74% | Conditions originating in the perinatal period | 98% |
| Diseases of the circulatory system | 73% | Pregnancy, childbirth and the puerperium | 91% |
| Certain infectious and parasitic diseases | 67% | Factors influencing health status | 91% |
| Diseases of the blood disorders & immune mechanism | 63% | Diseases of the eye and adnexa | 85% |
| Diseases of the genitourinary system | 60% | Neoplasms | 81% |

Figures 4 shows the inpatient prediction accuracy projected on patients' gender and age groups. No noticeable variations in predictions versus age or gender. This suggests the consistency of non-gender-age biases. The best accuracy was obtained for age group 60-70 mostly related to neoplasms and factors influencing health status categories.

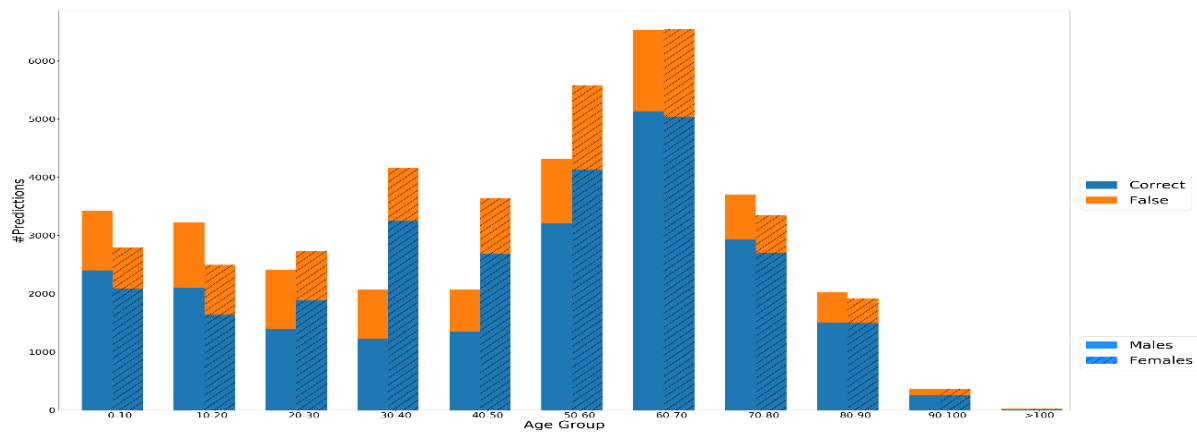

**Figure 4.** Inpatient predictions results shown over patients' gender across different age group

In Figure 5, the highest false predictions are for females of age groups 50-60 can be witnessed. Tracking back these cases, it was found that they are mostly cancer patients (breast). Prescriptions was found to vary for treating the main diagnosis and complications treated in follow-up visits, thus, with no historical link, it is hard to correctly predict many cases.

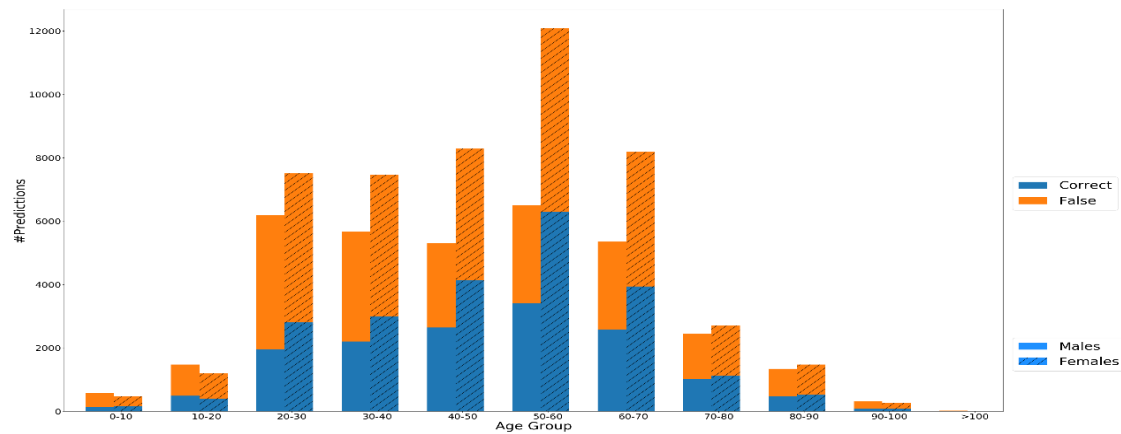

**Figure 5.** Outpatient predictions results shown over patients' gender across different age group

## 5. Conclusion

A novel collaborative residual learning based model was proposed to predict ICD10 codes exploiting knowledge effectively from prescriptions data only. Extensive experiments on two new real-life datasets showed significant improvements over baselines of varied NN architectures. The promising results suggests the possibility for further development by incorporating additional information such as medications' description, structure and dosage to discover salient diseases associations. Supplementary clinical data sources can also contribute to surge accuracy. We strongly recommend the adoption of one's model in real clinical coding practices to save time & cost, overcome understaffing, add consistency and allow for foreseeable coding.